\documentclass[12pt]{article}
\usepackage{graphicx}
\usepackage{amsmath}
\usepackage{amsfonts}
\usepackage{cite}
\begin{document}

\def\a{\alpha}
\def\b{\beta}
\def\c{\varepsilon}
\def\d{\delta}
\def\e{\epsilon}
\def\f{\phi}
\def\g{\gamma}
\def\h{\theta}
\def\k{\kappa}
\def\l{\lambda}
\def\m{\mu}
\def\n{\nu}
\def\p{\psi}
\def\q{\partial}
\def\r{\rho}
\def\s{\sigma}
\def\t{\tau}
\def\u{\upsilon}
\def\v{\varphi}
\def\w{\omega}
\def\x{\xi}
\def\y{\eta}
\def\z{\zeta}
\def\D{\Delta}
\def\G{\Gamma}
\def\H{\Theta}
\def\L{\Lambda}
\def\F{\Phi}
\def\P{\Psi}
\def\S{\Sigma}

\def\o{\over}
\def\beq{\begin{eqnarray}}
\def\eeq{\end{eqnarray}}
\newcommand{\gsim}{ \mathop{}_{\textstyle \sim}^{\textstyle >} }
\newcommand{\lsim}{ \mathop{}_{\textstyle \sim}^{\textstyle <} }
\newcommand{\vev}[1]{ \left\langle {#1} \right\rangle }
\newcommand{\bra}[1]{ \langle {#1} | }
\newcommand{\ket}[1]{ | {#1} \rangle }
\newcommand{\EV}{ {\rm eV} }
\newcommand{\KEV}{ {\rm keV} }
\newcommand{\MEV}{ {\rm MeV} }
\newcommand{\GEV}{ {\rm GeV} }
\newcommand{\TEV}{ {\rm TeV} }
\def\diag{\mathop{\rm diag}\nolimits}
\def\Spin{\mathop{\rm Spin}}
\def\SO{\mathop{\rm SO}}
\def\O{\mathop{\rm O}}
\def\SU{\mathop{\rm SU}}
\def\U{\mathop{\rm U}}
\def\Sp{\mathop{\rm Sp}}
\def\SL{\mathop{\rm SL}}
\def\tr{\mathop{\rm tr}}

\def\IJMP{Int.~J.~Mod.~Phys. }
\def\MPL{Mod.~Phys.~Lett. }
\def\NP{Nucl.~Phys. }
\def\PL{Phys.~Lett. }
\def\PR{Phys.~Rev. }
\def\PRL{Phys.~Rev.~Lett. }
\def\PTP{Prog.~Theor.~Phys. }
\def\ZP{Z.~Phys. }


\baselineskip 0.7cm
\begin{titlepage}

\begin{flushright}
IPMU13-218\\
ICRR-report-665-2013-14
\end{flushright}

\vskip 1.35cm
\begin{center}
{\bf Lower Bound on the Garvitino Mass
\\$m_{3/2}>O(100)$ TeV\\
in $R$-Symmetry Breaking New Inflation}

\vskip 1.2cm
Keisuke Harigaya$^1$, Masahiro Ibe$^{2,1}$ and Tsutomu T. Yanagida$^1$
\vskip 0.4cm
$^1${\it Kavli IPMU (WPI), TODIAS, University of Tokyo, Kashiwa 277-8583, Japan}\\
$^2${\it ICRR, University of Tokyo, Kashiwa 277-8582, Japan}
\vskip 1.5cm

\abstract{
In supersymmetric theories, the $R$ symmetry plays a unique role
in suppressing a constant term in the superpotential.
In  single chiral field models of spontaneous breaking of a discrete $R$ symmetry, 
an $R$-breaking field can be a good candidate for an inflaton
in new inflation models.
In this paper, we revisit the compatibility of the single-field $R$-breaking new
inflation model with the results of the Planck experiment.
As a result, we find that the model predicts a lower limit on the gravitino mass, 
$m_{3/2}>O(100)$\,TeV.
This lower limit is consistent with the
observed Higgs mass of $126$\,GeV
when the masses of the stops are of order the
gravitino mass scale.
}

\end{center}
\end{titlepage}

\setcounter{page}{2}

\section{Introduction}
In supersymmetric (SUSY) theories, the $R$ symmetry plays a unique role
in suppressing a constant term in the superpotential.
Without the $R$ symmetry, the constant term  is expected to be 
at the Planck scale, which requires a SUSY breaking scale to be the Planck scale to achieve the almost flat universe.
Thus, there is a strong case for the existence of a spontaneously broken
$R$-symmetry
if SUSY is the solution to the
hierarchy problem~\cite{MaianiLecture,Veltman:1980mj,Witten:1981nf,Kaul:1981wp}
between the weak scale and 
the  Planck scale or the scale of the Grand Unified Theory (GUT).

One caveat of the $R$ symmetry  is  that a generation of 
the appropriate vacuum expectation value (VEV)
of the superpotential 
requires a symmetry breaking field to have a Planck scale $A$-term VEV 
and a non-vanishing $F$ term VEV at the same time
if the symmetry is a continuous one~\cite{Dine:2009sw}.
This means that an $R$ symmetry breaking field is nothing but the Polonyi field
for the continuous $R$ symmetry.
Therefore, by taking the Polonyi problem~\cite{Coughlan:1983ci} seriously, 
the $R$ symmetry which suppresses the constant term of the superpotential 
should be a discrete one.

Interestingly, the simplest model of spontaneous discrete $R$ symmetry breaking consisting of
a single chiral field has a convex but a very flat potential
around the origin of the chiral field,%
\footnote{
Ref.~\cite{Kumekawa:1994gx} pointed out not only the
presence of the so-called $\eta$ problem in supergravity inflation models
but also the importance of the $R$ symmetry to have flat potentials
necessary for the inflation to occur.
}
which evokes a scalar potential used in new inflation models~\cite{Linde:1981mu,Albrecht:1982wi}.
In fact,  the simplest $R$-breaking model satisfies 
the slow-roll conditions in a wide parameter region, and hence, 
the $R$-breaking field is a good candidate for an inflaton~\cite{Kumekawa:1994gx,Izawa:1996dv,Izawa:1997df,Ibe:2006fs,Ibe:2006ck,Takahashi:2013cxa}.
It is also remarkable that the domain wall problem~\cite{Zeldovich:1974uw}
associated with the discrete $R$ symmetry breaking is automatically solved
when the $R$ symmetry breaking field plays a role of the inflaton.%
\footnote{
This situation is analogous 
to the original new inflation model~\cite{Linde:1981mu,Albrecht:1982wi},
where an inflaton is identified with a GUT breaking field and 
the monopole problem is solved.
}

We here emphasize that new inflation models tend to predict a small tensor fraction 
due to their small inflation scales~\cite{Lyth:1996im}.
This property is fairly supported by the upper limit on the tensor fraction
of cosmic perturbations set by the recent
observations of the cosmic microwave background (CMB) ~\cite{Hinshaw:2012aka,Ade:2013zuv,Ade:2013uln}.%
\footnote{
Simple large field inflation models such as the chaotic inflation models
with a quadratic or a quartic potential~\cite{Linde:1983gd} are, on the other hand, 
now slightly disfavored  at least by $1\sigma$ level,
which requires some extensions~\cite{Kallosh:2010ug,Takahashi:2010ky,Harigaya:2012pg,Croon:2013ana,Nakayama:2013jka}.
}

In this paper, we further investigate the compatibility of the $R$-breaking new
inflation model with the results of the Planck
experiment~\cite{Ade:2013zuv,Ade:2013uln}.
As we will see,  the $R$-breaking new inflation model is
consistent with all cosmological constraints and observations
in a wide parameter region.
Furthermore, the model predicts a lower bound on the gravitino mass, 
$m_{3/2}>O(100)$\,TeV.
This lower limit on the gravitino mass is consistent with the
observed Higgs mass of $126$\,GeV~\cite{Aad:2012tfa,Chatrchyan:2012ufa}
in a class of models in which the masses of the stops are of order the
gravitino mass~\cite{Okada:1990vk,Ellis:1990nz,Haber:1990aw}.
We also show that 
the baryon asymmetry of the universe as well as the observed dark matter density 
can be consistently explained along with the $R$-breaking new inflation model.

\section{Brief review on the $R$-breaking new inflation model}
Let us begin with the simplest model of spontaneous discrete $Z_{NR}$ 
symmetry breaking consisting of a single chiral field $\phi$~\cite{Kumekawa:1994gx,Izawa:1996dv}.
Here, we assume that $\phi$ is a singlet except for the $R$ symmetry with an $R$ charge $2$. 
Assuming $N=2n$,  the superpotential of $\phi$ is given by,
\label{eq:super}
\begin{eqnarray}
W=v^2\phi -\frac{g}{n+1}\phi^{n+1} + \cdots,
\end{eqnarray}
where the ellipses represent higher power terms of $\phi$.
We neglect them throughout this paper, since
we are interested in the region with $|\phi|\ll1$.
The size of the coupling constant $g$ will be discussed later. 
Here and hereafter, we take the unit of the reduced Planck scale
$M_{PL}\simeq 2.4\times 10^{18}$ GeV being unity.
The parameters $v^2$ and $g$ are taken real and positive without loss of generality.%
\footnote{
In order for the gravitino mass to be far smaller than the Planck scale,
$v^2$ must be suppressed.
The suppression can be explained, for example, by assuming an $U(1)_R$ symmetry under
which $\phi$ has a charge of $2/(n+1)$, and the $U(1)_R$ symmetry
be dynamically broken by a condensation of a (composite) chiral field with an $U(1)_R$ charge
of $2-2(n+1)$~\cite{Izawa:1996dv}.
}
At supersymmetric vacua, the $Z_{2nR}$ symmetry is spontaneously broken down to
the $Z_{2R}$ symmetry
by the VEV of $\phi$,
\begin{eqnarray}
 \vev{\phi}\simeq \left(\frac{v^2}{g}\right)^{1/n}\times e^{2\pi i
  m/n},
  ~~~~(m = 0,1,\cdots,n-1)
\end{eqnarray}
which leads to the VEV of the superpotential,
\begin{eqnarray}
\vev{W}\simeq \frac{n}{n+1}v^2 \left(\frac{v^2}{g}\right)^{1/n}
  e^{2\pi i m/n} .
\end{eqnarray}

As we emphasized in the introduction, 
the scalar potential of this model is convex but very flat around $\phi \sim 0$.
Thus, if the initial field value of $\phi$ is set close to its origin by, for example,
a positive Hubble induced mass term of pre-inflation~\cite{Izawa:1997df} 
and the slow-roll conditions are satisfied, $\phi$ automatically brings about the inflation.
Therefore, the simplest model of  discrete $R$-symmetry breaking 
is equipped with necessary structures as a model of new inflation.

Now, let us discuss details of the new inflation model.
For that purpose, let us note that the K\"ahler potential of $\phi$ is given by
\begin{eqnarray}
\label{eq:Kahler}
 K = \phi \phi^\dag + \frac{1}{4}k \left(\phi \phi^\dag \right)^2 + \cdots,
\end{eqnarray}
where the ellipses denote higher power terms of $\phi$, whose
contributions to the dynamics of $\phi$ are negligible again.
The parameter $k$ is at most of order unity, and we assume
$k>0$ so that $\phi = 0$ is a local maximum (see below).
From Eqs.~(\ref{eq:super}) and (\ref{eq:Kahler}),
the scalar potential of the scalar component of $\phi$ is given by
\begin{eqnarray}
 V(\phi) &=& |v^2-g\phi^n|^2 - k v^4 |\phi|^2 + \cdots\nonumber\\
&=& v^4 -\left(g v^2\phi^n + {\rm h.c.}\right) - kv^4|\phi|^2 \cdots \ .
\end{eqnarray}
In terms of the radial and the angular components of $\phi$, 
 $\phi = \varphi e^{i\theta}/\sqrt{2}$, 
the scalar potential is rewritten as,
\begin{eqnarray}
 V(\varphi,\theta) = v^4 -\frac{k}{2}v^4 \varphi^2
  -\frac{g}{2^{n/2-1}}v^2 \varphi^n {\rm cos}\left(n\theta\right) +\cdots.
\end{eqnarray}
It can be seen that for a given $\varphi>0$, the minimum of the
potential is provided by $\theta = 2\pi l/n~(l=0,1,\cdots,n-1)$.
In the following, the radial component $\varphi$ plays 
a role of the inflaton in new inflation.

As we have mentioned, we assume that the initial condition
of $\varphi$ is close to $0$, i.e. $|\varphi| \ll 1$.
We further  suppose that the initial condition of the angular
direction $\theta$ is given by
$\theta = 0~({\rm mod}~2\pi/n)$ for the time being.
Since $\theta = 0~({\rm mod}~2\pi/n)$ is the minimum of the potential along the angular
direction, $\theta = 0~({\rm mod}~2\pi/n)$ is kept during the inflation.
Along the inflaton trajectory, the first and the second slow-roll parameters are given by
\begin{eqnarray}
 \epsilon &\equiv& \frac{1}{2}\left(\frac{\partial V/\partial \varphi}{V}\right)^2 =
\frac{1}{2}\left(k \varphi+
\frac{ng}{2^{n/2-1}}\frac{\varphi^{n-1}}{v^2}\right)^2, \nonumber\\
 \eta &\equiv& \frac{\partial^2 V/\partial \varphi^2}{V} =
- k -
\frac{n(n-1)g}{2^{n/2-1}}\frac{\varphi^{n-2}}{v^2}.
\end{eqnarray}
Thus, the slow-roll conditions can be actually satisfied for $|\varphi|\ll 1$
as long as $k\ll 1$.

By assuming $|k| \ll 1$,  the inflation lasts until the inflaton
reaches to 
\begin{eqnarray}
  \varphi_{\rm end} =  \left(\frac{2^{(n-2)/2}v^2}{n(n-1)g}\right)^{1/(n-2)}\ ,
\end{eqnarray}
at which the slow-roll conditions are violated, $|\eta| \simeq 1$.
It should be noted that there is an one-to-one correspondence 
between the number of $e$-foldings $N_e$ and the field value of $\varphi$ during
the inflation via 
\begin{eqnarray}
\label{eq:efold}
 N_e (\varphi) = \int_{\varphi_{\rm end}}^\varphi \frac{V}{\partial
  V/\partial \varphi}{\rm d}\varphi \ .
\end{eqnarray}
Thus, by taking  the inverse of Eq.~(\ref{eq:efold}), we obtain
\begin{eqnarray}
 \varphi^{n-2}(N_e) = \frac{2^{(n-2)/2}kv^2}{ng}\left(
e^{k(n-2)N_e}-1 + k(n-1)e^{k(n-2)N_e}
\right)^{-1}.
\end{eqnarray}

\begin{figure}[tb]
 \begin{center}
  \includegraphics[width=0.5\linewidth]{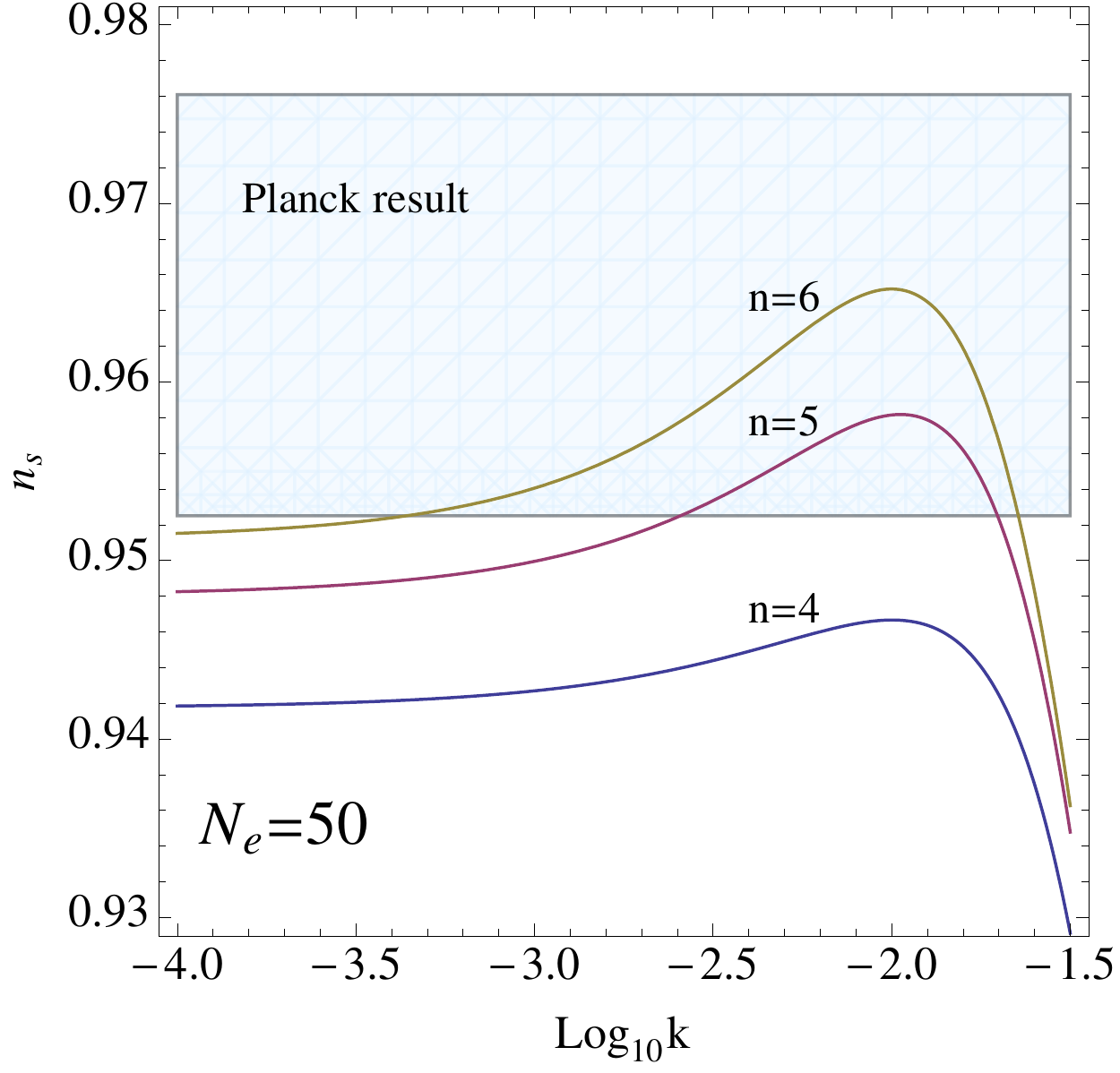}
 \end{center}
\caption{The spectral index of the curvature perturbation $n_s$ for $n=4,5,6$
 with $N_e=50$.
A colored region show the 95\% C.L. favored region by the Planck
experiment, $n_s = 0.9643\pm 0.012$~\cite{Ade:2013uln}.
}
\label{fig:tilt}
\end{figure}

\begin{figure}[tb]
 \begin{center}
  \includegraphics[width=0.5\linewidth]{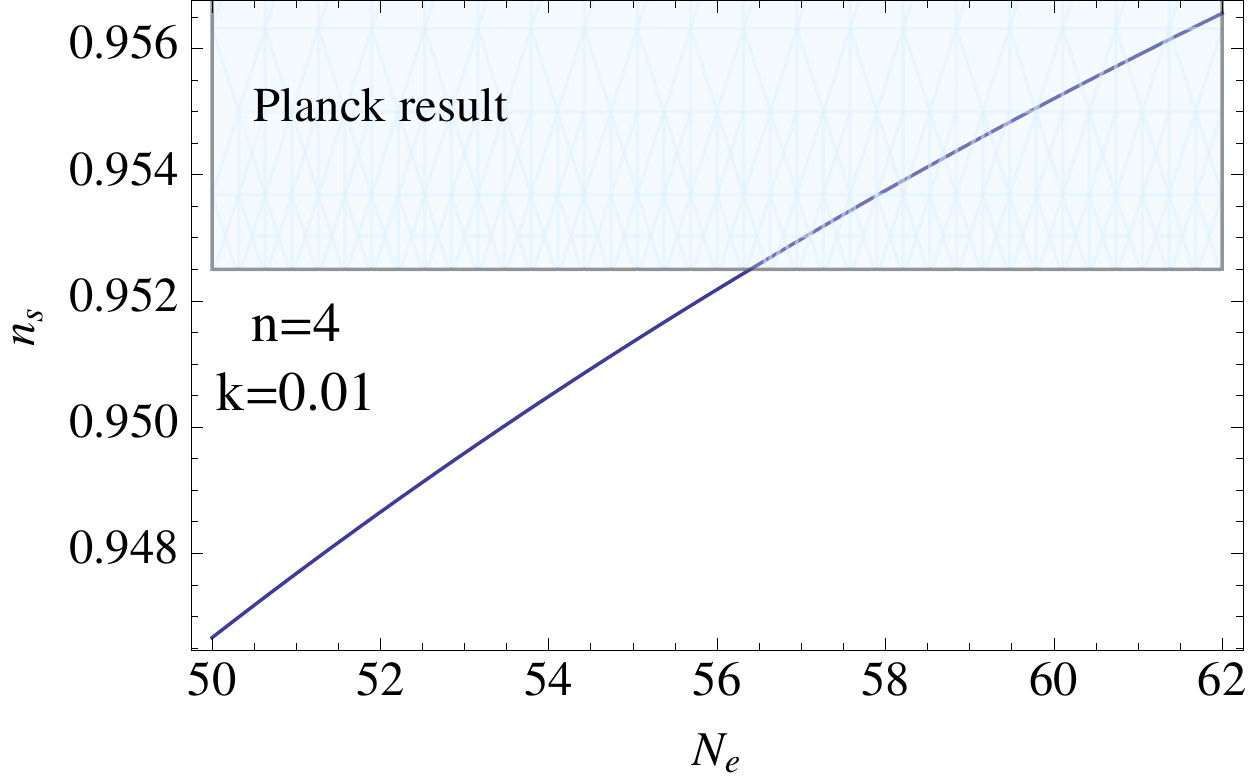}
 \end{center}
\caption{The spectral index of the curvature perturbation $n_s$ for $n=4$
 with various $N_e$.
The colored region show the 95\% C.L. limit from the Planck
experiment, $n_s = 0.9643\pm 0.012$~\cite{Ade:2013uln}.
}
\label{fig:tilt2}
\end{figure}

In order to compare model predictions with 
CMB observations, 
let us calculate the properties of the curvature perturbation.
The spectrum of the curvature perturbation ${\cal P}_\zeta$
and its spectral index $n_s$
are given by
\begin{eqnarray}
\label{eq:Pzeta}
 {\cal P}_\zeta &=& \frac{1}{24\pi^2}\frac{V}{\epsilon}=
\frac{1}{24\pi^2}\left(
n^2g^2 k^{-2(n-1)}v^{4(n-3)}\left(e^{k(n-2)N_e}-1\right)^{2(n-1)}
\right)^{\frac{1}{n-2}}e^{-2k(n-2)N_e},
 \\
n_s &=& 1-6\epsilon+2\eta= 1-2k\left(
1 + \frac{n-1}{\left(1+k\left(n-1\right)\right)e^{k(n-2)N_e}-1}
\right),
\end{eqnarray}
respectively.
In Fig.~\ref{fig:tilt}, we show the prediction on the spectral index
for $n=4,5,6$ and $N_e=50$. The colored region shows a region favored by the Planck 
experiment, i.e. $n_s = 0.9643\pm 0.012$~\cite{Ade:2013uln}
for the pivot scale $k_*=0.002~{\rm Mpc}^{-1}$ at 95\%C.L.
It can be seen that the model with $n\leq4$ is disfavored by the Planck
experiment for $N_e =50$.
For $n=5$, $k\sim 10^{-2}$ is favored.
In Fig.~\ref{fig:tilt2}, we show the $N_e$ dependence of the spectral index for $n=4$.
The figure shows that
the model with $n=4$ is still consistent with the
Planck experiment for $N_e\gsim 56$.%
\footnote{
In Ref.~\cite{Takahashi:2013cxa}, it is pointed out that the model with $n=4$ is also
consistent with the Planck experiment if there are a small constant term
in the superpotential beside the one from the condensation of $\phi$.
Since we assume that the $R$ symmetry is broken only by the condensation
of $\phi$, that solution is not applicable.
}
We will discuss impacts of the observed spectral index to the gravitino mass
in the next section.

\begin{figure}[tb]
 \begin{center}
  \includegraphics[width=0.4\linewidth]{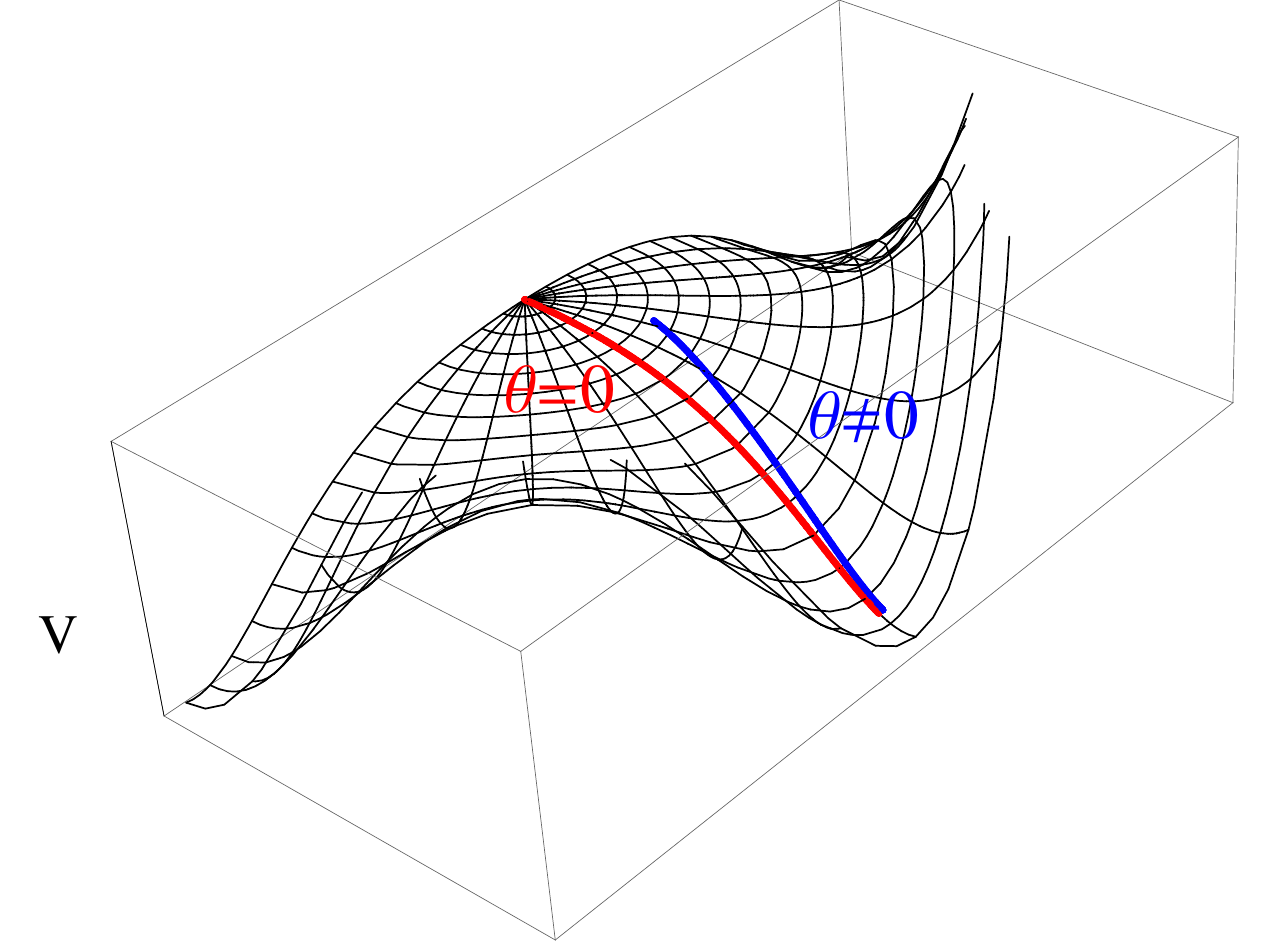}
 \end{center}
\caption{A schematic picture for the scalar potential of $\phi$.
The two lines show the trajectories of the inflaton with 
angular initial condition with either $\theta = 0$ or $\theta \neq 0$.
The later trajectory feels steeper potential, and hence, the spectral index 
becomes more red-tilted.
}
\label{fig:potential}
\end{figure}

Before closing this section, let us discuss more general initial conditions
for the inflaton field, $\theta \neq 0~({\rm mod}~2\pi/n)$.
In particular, we are interested in how the spectral index is affected, since $n=4$ is
severely constrained for $\theta = 0~({\rm mod}~2\pi/n)$ by the Planck results.
In Fig.~\ref{fig:potential}, we show a schematic picture of the
shape of the inflaton potential for $n=4$.
For a better presentation, we show only the region with ${\rm Re}(\phi)>0$.
For a fixed number of e-foldings, a non-zero
angle $\theta$ leads to a larger corresponding field value for $\varphi$. As a result,
the curvature of the inflaton trajectory becomes negatively larger, and
the spectral index becomes more red-tilted.
Therefore, even if we consider the initial condition with $\theta \neq 0~({\rm mod}~2\pi/n)$, the model with $n=4$ is still disfavored unless $N_e$ is large.

\section{Lower bound on the gravitino mass}
In this section, we put a lower bound on the gravitino mass $m_{3/2}$ in the
$R$-breaking new inflation models based on the results obtained in the
previous section.
From Eq.~(\ref{eq:Pzeta}), the parameter $v^2$ is
expressed by the curvature perturbation, ${\cal P}_\zeta \simeq 2.2\times10^{-9}$~\cite{Ade:2013uln}, as
\begin{eqnarray}
\label{eq:v}
 v^2 &=& \left(
\left(24\pi^2{\cal P}_\zeta\right)^{n-2}
\left(ng\right)^{-2}
\left(\frac{k}{e^{k(n-2)N_e}-1}\right)^{2(n-1)}
e^{2k(n-2)^2N_e}
\right)^{\frac{1}{2(n-3)}}\ ,
\end{eqnarray}
which leads to
\begin{eqnarray}
\label{eq:v}
 v &\simeq&
\begin{cases}
9.0\times10^{11}~{\rm GeV}~g^{-1/2}&(n=4,k=0.01,N_e=56),\\
6.2\times10^{13}~{\rm GeV}~g^{-1/4}&(n=5,k=0.01,N_e=50),\\
2.5\times10^{14}~{\rm GeV}~g^{-1/6}&(n=6,k=0.01,N_e=50).
\end{cases}
\end{eqnarray}
It should be noted that $v$ does not depend on $k$ significantly.
As a result,  the gravitino mass $m_{3/2}$ is given by
\begin{eqnarray}
\label{eq:m32}
 m_{3/2} &=& \frac{ng}{n+1}
 \left(\frac{v^2}{g}\right)^{\frac{n+1}{n}}\nonumber\\
&\simeq&
\begin{cases}
1.6\times10^{2}~{\rm GeV}~g^{-3/2}&(n=4,k=0.01,N_e=56),\\
2.0\times10^{7}~{\rm GeV}~g^{-4/5}&(n=5,k=0.01,N_e=50),\\
1.1\times10^{9}~{\rm GeV}~g^{-5/9}&(n=6,k=0.01,N_e=50).
\end{cases}
\end{eqnarray}

\begin{figure}[tb]
 \begin{center}
  \includegraphics[width=0.5\linewidth]{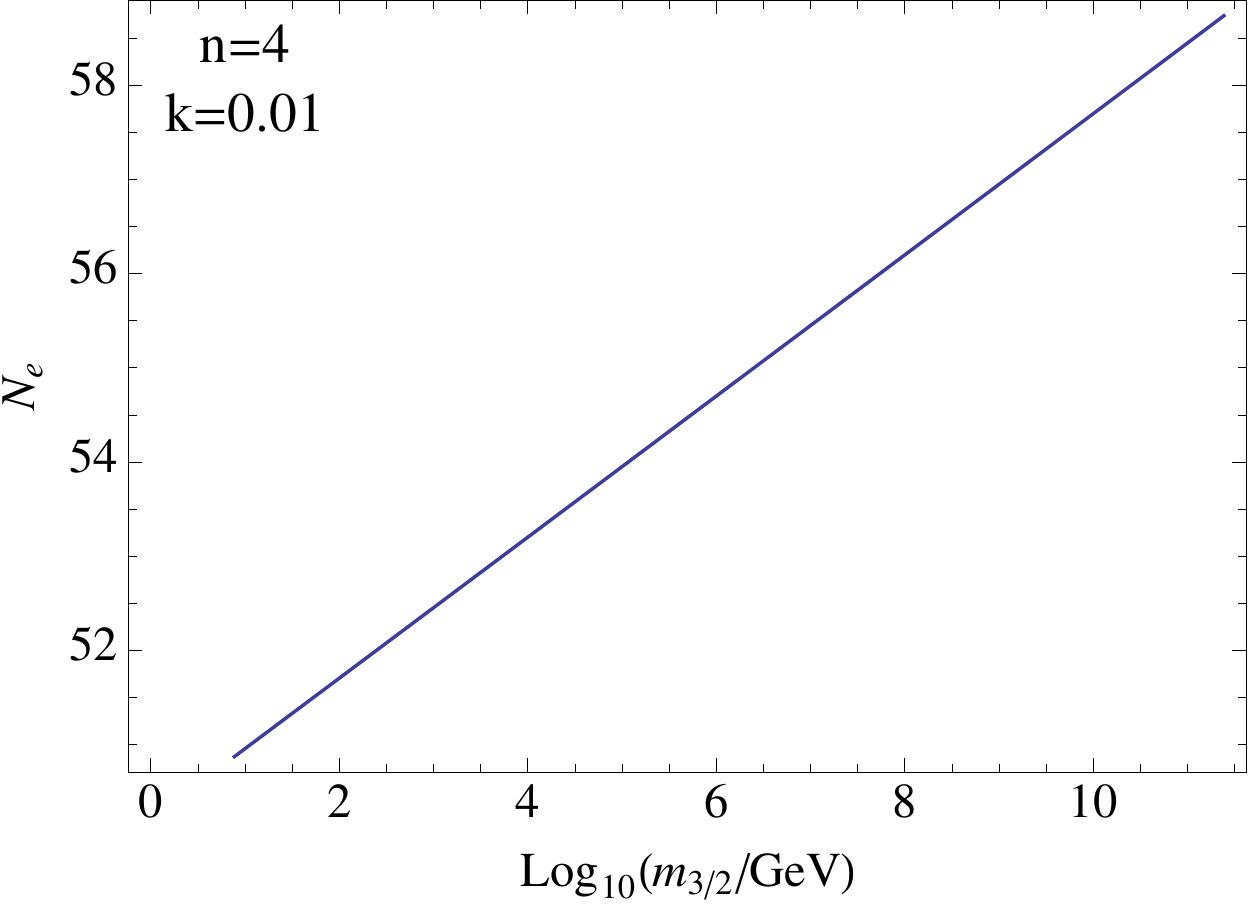}
 \end{center}
\caption{A relation between $m_{3/2}$ and $N_e$ for $n=4$.
}
\label{fig:m32Ne}
\end{figure}

As we have shown in the previous section, the model with $n=4$ is
consistent with the Planck experiment only if $N_e\gsim 56$.
This requires a very large $v^2$, which in turn puts a lower bound on the
gravitino mass.
To see this, let us remind ourselves that $N_e$ is given by the inflation scale as~\cite{Liddle:1993fq}
\begin{eqnarray}
\label{eq:Ne}
 N_e = 52-{\rm ln}\left(\frac{10^{12}~{\rm
			      GeV}}{v}\right)\ ,
\end{eqnarray}
for the pivot scale $k_*=0.002~{\rm Mpc}^{-1}$.
Here, 
we have assumed an instantaneous reheating after the inflation, which brings
about the largest $N_e$ for a fixed inflation scale.
From Eqs.~(\ref{eq:v}), (\ref{eq:m32}) and (\ref{eq:Ne}), we obtain a
relation between $m_{3/2}$ and $N_e$,
which is shown in Fig.~\ref{fig:m32Ne}.
From the figure and the constraint $N_e\gsim 56$, we obtain a lower
bound on the gravitino mass, $m_{3/2}>{\cal O}(10^8)$ GeV.

\begin{figure}[tb]
 \begin{center}
  \includegraphics[width=0.5\linewidth]{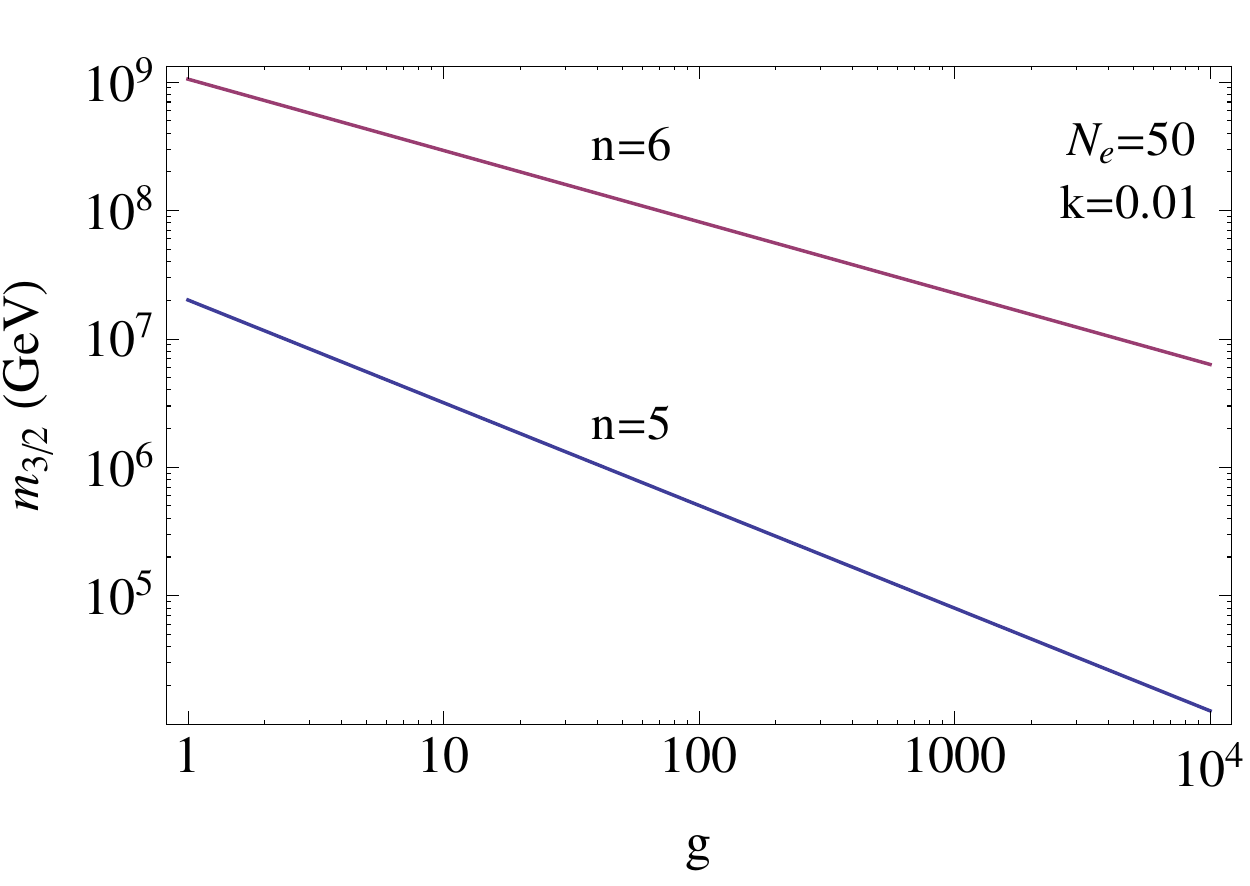}
 \end{center}
\caption{The gravitino mass for $n=5,6$ with $N_e=50$, $k=0.01$.
}
\label{fig:m32}
\end{figure}

Next, let us discuss the model with $n>4$.
In Fig.~\ref{fig:m32}, we show the gravitino mass for $n=5,6$ with
$N_e=50$, $k=0.01$.
In can be seen that the larger and smaller $n$ and $g$ are, the larger
the gravitino mass is.
Hence, we can derive a lower bound on $m_{3/2}$ from a upper bound on
$g$ for the model with $n=5$.

It should be noted that there is an upper bound on $g$ from the unitarity limit,
which can be extracted by considering the leading radiative 
correction to the K\"ahler potential due to the coupling $g$,
\begin{eqnarray}
 \delta K \simeq \frac{5!}{(16\pi^2)^4}g^2M_*^6 \phi \phi^\dag,
\end{eqnarray}
where $M_*$ is the cutoff of the loop integration.
By requiring the unitarity up to the Planck scale, i.e. $M_* \simeq M_{\rm PL}$,
the unitarity limit, $|\delta K|\lsim \phi\phi^\dag$,
leads to an upper bound on $g$,%
\footnote{
This requirement based on $M_* =1 $ is equivalent to the Born unitarity up to the Planck scale.
}
\begin{eqnarray}
\label{eq:uplimg}
 g \lsim (16\pi^2)^2/\sqrt{5!}\simeq 2000.
\end{eqnarray}
By substituting this upper limit into Eqs.~(\ref{eq:m32}) and (\ref{eq:uplimg}),
we obtain a lower bound on the
gravitino mass, $m_{3/2}\gsim 100~{\rm TeV}$
for $n > 4$.

In summary, 
we find that the lower bound on the gravitino mass; 
\begin{eqnarray}
\label{eq:gravitino mass}
 m_{3/2}\gsim 100~{\rm TeV},
\end{eqnarray}
in the $R$-breaking new inflation model.
For $n =4$, the (much higher) lower limit on the gravitino mass is obtained 
to achieve the observed spectral index, while the milder limit for $n>4$
is obtained from the size of the curvature perturbation.
As stressed in the introduction, this lower bound is consistent with the
observed Higgs mass of $125$ GeV~\cite{Okada:1990vk,Ellis:1990nz,Haber:1990aw}.

\section{Baryon asymmetry and dark matter density}
In this section, we argue that the baryon asymmetry as well as the dark matter density in the present universe
can be explained consistently with the $R$-breaking inflation model.
In the following, we concentrate on the model with $n=5$, $k\simeq 0.01$
and $N_e=50$.

\subsection{Baryon asymmetry}
\subsubsection*{Thermal leptogenesis}
Let us first discuss whether the thermal leptogenesis~\cite{Fukugita:1986hr} can be
achieved in the $R$-breaking new inflation model, that is, whether a
reheating temperature $T_R$ can be high enough, $T_R\gsim 10^9$ GeV~\cite{Buchmuller:2004nz}.

First, let us consider an inflaton decay via Planck-suppressed dimension five
interactions%
\footnote{
For example, a K\"ahler interaction $K= \lambda\phi^\dag QQ$, where
$Q$ is some chiral field lighter than the inflaton, provides such decay channel.}
in which 
the decay width of the inflaton $\Gamma_{\phi,{\rm dim}-5}$ is as large
as $m_{\phi}^3$,
where $m_{\phi}$ is the inflaton mass around the
vacuum,
\begin{eqnarray}
 m_\phi = 5 g \left(\frac{v^2}{g}\right)^{4/5} \simeq 1.4\times 10^{11}~{\rm
  GeV} \left(\frac{g}{1000}\right)^{-1/5}.
\end{eqnarray}
In this case,
a reheating temperature $T_R$ is as large as
\begin{eqnarray}
\label{eq:TR_dim5}
 T_R\sim \sqrt{\Gamma_{\phi,{\rm dim}-5}} \sim 10^7~{\rm
  GeV}\left(\frac{g}{1000}\right)^{-3/10}\ll 10^9~{\rm GeV}.
\end{eqnarray}
Therefore, for a successful thermal leptogenesis, we are lead to
introduce unsuppressed interactions.%
\footnote{
If the dimension five interaction saturates the unitarity bound,
$\lambda\sim 4\pi$, $T_R$ is
as large as $10^8$ GeV.
When
the right-handed neutrinos have a non-hierarchical mass spectrum and
the neutrino Yukawa matrix is rather tuned, the
thermal leptogenesis is possible~\cite{Flanz:1996fb,Pilaftsis:1997jf,Blanchet:2008pw}.
}

In order to enhance the decay rate of the inflaton,
let us consider a superpotential
\begin{eqnarray}
\label{eq:super_decay}
 W = \frac{y}{2\ell} \phi^\ell QQ,
\end{eqnarray}
where $Q$ is some chiral field lighter than the inflaton and 
$y$ is a coupling constant.
Due to large $\vev{\phi}$,
\begin{eqnarray}
 \vev{\phi}=\left(\frac{v^2}{g}\right)^{1/5}\simeq 2\times 10^{-3}\left(\frac{g}{1000}\right)^{-3/10},
\end{eqnarray}
the decay of the inflaton by this interaction
is effective even if $\ell>1$.
The decay width of $\phi$ by this operator is given by
\begin{eqnarray}
 \Gamma_\phi = \frac{1}{8\pi} y^2 |\vev{\phi}|^{2\ell-2}m_{\phi}
  =\frac{\ell^2}{8\pi}\frac{m_Q^2}{|\vev{\phi}|^2} m_{\phi},
\end{eqnarray}
where $m_Q$ is the mass of $Q$.
A reheating temperature is given by
\begin{eqnarray}
\label{eq:TR}
 T_R \simeq \left(\frac{90}{\pi^2 g_*}\right)^{1/4}\sqrt{\Gamma_\phi}=
1.8\times~10^9
{\rm GeV}
 \left(\frac{m_Q}{5\times10^{10}{\rm GeV}}\right)
 \left(\frac{\ell}{3}\right)
 \left(\frac{g}{1000}\right)^{1/5}
 \left(\frac{g_*}{200}\right)^{-1/4},
\end{eqnarray}
where $g_*$ is the effective degree of freedom of the radiations.
It can be seen that the thermal leptogenesis is marginally possible.

In the mentioned above reheating scenario, we have introduced
a matter field $Q$.
Note that we cannot identify $Q$ with the minimal supersymmetric
standard model (MSSM) higgs doublets, since a Dirac
mass term of the MSSM higgs doublets, the so-called $\mu$ term, should be
as small as the gravitino mass, and hence a reheating temperature is
not high enough (see Eq.~(\ref{eq:TR})).

An interesting idea is to identify $Q$ with the right-handed
neutrinos, $N_i~(i=1,2,3)$~\cite{Kumekawa:1994gx}. 
In this case, the masses of the right-handed neutrinos,
which should be far smaller than the Planck scale in order to obtain the
observed masses of the left-handed neutrinos by the seesaw mechanism~\cite{seesaw},
are controlled by
the $Z_{2nR}$ symmetry rather than the $B-L$ symmetry.

For example, let us arrange the right-handed neutrinos by their masses; $m_{N_1}\leq
m_{N_2}\leq m_{N_3}$. The inflaton decays mostly into the heaviest
right-handed neutrino as long as the decay is kinematically allowed,
that is, $2m_{N_i}<m_{\phi}$. If the inflaton decays mostly
into $N_2$ or $N_3$ and the resulting reheating temperature is larger enough
than $m_{N_1}$, the thermal leptogenesis is marginally possible.

\subsubsection*{Non-thermal leptogenesis}
We have shown that the thermal leptogenesis is marginally possible in
the $R$-breaking new inflation model with $n=5$.
Interestingly, when we identify $Q$ with the right-handed neutrinos, 
a possibility of the non-thermal
leptogenesis scenario~\cite{Ibe:2006fs} is also opened.%
\footnote{
If we introduce a K\"ahler interaction $K=\phi^\dag NN$ instead of the
superpotential given by Eq.~(\ref{eq:super_decay}), a reheating temperature is as large as $10^7$ GeV (Eq.~(\ref{eq:TR_dim5})) and
the non-thermal leptogenesis is possible.
In this case, the right-handed neutrinos has an $R$ charge
of one, and the masses of the right-handed neutrinos are in general of order the Planck scale. In order
to obtain $m_N<m_\phi$ as well as the observed masses of the left-handed
neutrinos, some tunings are necessary.
If we further assume that the scale $v^2$ is given by a breaking of
some charged field, the masses of the right-handed neutrinos are also
given by breaking of the charged field and hence is naturally small.
}
There, the inflaton decays into right-handed
neutrinos and the non-equilibrium decay of the right-handed neutrinos with
a $CP$ violation generates lepton numbers.

For simplicity, let us assume that the inflaton decays mostly into the lightest
right-handed neutrino $N_1$.
The entropy yield of the baryon number is given by~\cite{Ibe:2005jf}
\begin{eqnarray}
\eta_B\equiv \frac{n_B}{s}= 9\times 10^{-11} \left(\frac{T_R}{10^6{\rm
  GeV}}\right)
  \left(
  \frac{2m_{N_1}}{m_\phi}\right)
 \left( \frac{m_{\nu3}}{0.05{\rm eV}}\right)
 \frac{1}{{\rm
  sin}^2\beta}\delta_{\rm eff},
\end{eqnarray}
where $m_{\nu3}$ is the mass of the heaviest
left-handed neutrino, and $\beta$ is defined by the
vacuum expectation values of the up-type and down-type higgs doublets, $H_u$
and $H_d$, as
${\rm tan}\beta = \vev{H_u}/\vev{H_d}$.
$\delta_{\rm eff}$ represents a degree of the $CP$ violation, which is
given by the Yukawa couplings of the right-handed neutrinos, and
expected be of order one.
Compared with the observed value, $\eta_{B-{\rm obs}} \simeq 8.5\times
10^{-11}$~\cite{Ade:2013zuv},
an appropriate baryon asymmetry can be generated in the non-thermal
leptogenesis scenario.

\subsection{Dark matter density}
In the MSSM, there is a
candidate for dark matter, the lightest supersymmetric particle
(LSP).
Here, we assume pure gravity mediation models/minimal split SUSY models~\cite{Ibe:2006de,minimalSplit}, in which the gaugino
masses are generated only by one-loop effects and hence smaller in comparison
with the gravitino, higgsino and sfermion masses, and the wino is the LSP.
The wino mass $M_2$ is given by~\cite{Giudice:1998xp}
\begin{eqnarray}
\label{eq:AMSB}
 M_2 = \frac{g_2^2}{16\pi^2}\left(m_{3/2}+L\right),
\end{eqnarray}
where $g_2$ is the $SU(2)$ gauge coupling constant. 
The first term originates from an anomaly mediated effect~\cite{Dine:1992yw,Randall:1998uk,Giudice:1998xp},
while the second term, $L$, parametrizes a higgsino threshold
correction.%
\footnote{
If there is a vector-like matter in addition to the MSSM fields, the
gaugino masses receive a one-loop correction further~\cite{Nelson:2002sa,Nakayama:2013uta}.
For a comprehensive discussion on the phenomenology of the gauginos in
that case, see Ref.~\cite{Harigaya:2013asa}.
}
As shown in Ref.\,\cite{Ibe:2006de}, $L$ is expected to be of order the gravitino mass in pure gravity mediation
models/minimal split SUSY models.

There are three sources for wino productions,
a thermal wino relic, non-thermal production of gravitinos from a thermal
bath, and gravitino production from the inflaton decay.
We explain them in the following.

\subsubsection*{Thermal wino relic}
Since the wino has an $SU(2)$ gauge interaction, it is in a thermal
equilibrium in the early universe. As the temperature of the universe
decreases, the wino abundance freezes out
and remains as a dark matter since the wino is the LSP.
This is nothing but the
conventional WIMP scenario.
In order for the thermal abundance not to excess the observed cold dark
matter value, $\Omega_c h^2=0.1196\pm0.0031$~\cite{Ade:2013zuv}, it is required that~\cite{Hisano:2006nn}
\begin{eqnarray}
\label{eq:M2th}
 M_2 \lsim 3~{\rm TeV}.
\end{eqnarray}

\subsubsection*{Gravitino scattered from thermal bath}
Since the gravitino interacts with another light fields only through
Planck-suppressed interactions, once it is scattered from a thermal
bath, it does not interact with the thermal bath again,
and eventually decays into the wino.
A contribution to the wino abundance from this process is given by~\cite{Kawasaki:1994af,Gherghetta:1999sw,Ibe:2004tg}
\begin{eqnarray}
\label{eq:M2sc}
 \Omega_{\rm wino,sc}h^2 \simeq 0.12\left( \frac{M_2}{200~{\rm
  GeV}}\right)\left(\frac{T_R}{10^{10}~{\rm GeV}}\right). 
\end{eqnarray}

\subsubsection*{Gravitino from inflaton decay}
After SUSY breaking, 
there is no remaining symmetry which prevents a mixing between the
inflaton field and the SUSY breaking field at the vacuum.
This effect induces an inflaton decay into gravitinos~\cite{Kawasaki:2006gs,Asaka:2006bv,Dine:2006ii,Endo:2006tf,Kawasaki:2006hm,Endo:2006qk,Endo:2007ih,Endo:2007sz}, 
which provides another source of  non-thermal wino dark matter.

As an example, let us take the
following effective superpotential for the SUSY breaking field $Z$,
\begin{eqnarray}
W_{\rm eff}= \Lambda^2 Z,
\end{eqnarray}
where $\Lambda^2$ is a SUSY breaking scale, which should satisfy
$\Lambda^2=\sqrt{3}m_{3/2}$ in our flat universe.%
\footnote{We have assumed that $|\vev{Z}|\ll1$ to avoid the Polonyi problem.}
By calculating the scalar potential of the scalar components of $Z$ and
$\delta\phi\equiv \phi-\vev{\phi}$ including supergravity effects, we obtain a mixing term,
\begin{eqnarray}
\label{eq:mixing}
 V_{\rm mix} = \sqrt{3}(1-b) m_{\phi} \vev{\phi} m_{3/2}\delta\phi Z^\dag +
  {\rm h.c.},
\end{eqnarray}
where $b$ is a coupling constant in the K\"ahler potential,
$K\supset b ZZ^\dag \phi \phi^\dag$. A mixing angle $\epsilon$ between the scalar
components of $Z$ and $\delta\phi$ is given by
\begin{eqnarray}
 \epsilon = \sqrt{3}(1-b) m_{\phi} \vev{\phi} m_{3/2}/m_Z^2,
\end{eqnarray}
where $m_Z$ is the mass of the SUSY breaking field.
Here it is assumed that $m_Z\gg m_{\phi}$, which is the case with
typical dynamical SUSY breaking models.%
\footnote{If not, an inflaton decay into gravitinos is suppressed~\cite{Dine:2006ii,Endo:2006tf}.
An inflaton decay into SUSY breaking sector fields, which are expected
to exist in general dynamical SUSY breaking models, can be also
suppressed by separating the dynamical scale and the mass of $Z$,
$m_Z\ll \Lambda$~\cite{Nakayama:2012hy}.
}

A coupling between the scalar component of $Z$ and its fermionic
component $\psi$, the goldstino, is provided by the following K\"ahler potential which
gives a mass to the scalar component of the SUSY breaking field~\cite{Dine:2006ii,Nakayama:2012hy},
\begin{eqnarray}
\label{eq:mZ Kahler}
 K\supset -\frac{m_Z^2}{12m_{3/2}^2}ZZ^\dag Z Z^\dag.
\end{eqnarray}
The $D$ term of Eq.~(\ref{eq:mZ Kahler}) yields
\begin{eqnarray}
 {\cal L}\supset -\frac{\sqrt{3}}{6}\frac{m_Z^2}{m_{3/2}}Z^\dag \psi
  \psi + {\rm h.c.} 
\end{eqnarray}

From Eqs~(\ref{eq:mixing}) and (\ref{eq:mZ Kahler}), the decay rate is given by
\begin{eqnarray}
\Gamma_{3/2}\equiv \Gamma_{\phi\rightarrow 2\psi_{3/2}}  \simeq \Gamma_{Z\rightarrow
  2\psi,m_Z=m_{\phi}}|\epsilon|^2 = \frac{(b-1)^2}{32\pi} m_{\phi}^3 \vev{\phi}^2.
\end{eqnarray}
The entropy yield of the gravitino after the inflaton decay, $Y_{3/2}$,
is estimated as
\begin{eqnarray}
 Y_{3/2} = 2\times \frac{\Gamma_{3/2}}{\Gamma_{\rm tot}} \frac{3 T_R}{4
  m_{\phi}} = \frac{3}{2} \sqrt{\frac{\pi^2 g_*}{90}}\frac{\Gamma_{3/2}}{m_{\phi}T_R},
\end{eqnarray}
where $\Gamma_{\rm tot}$ is a total decay width of the inflaton.
The wino abundance is given by
\begin{eqnarray}
\label{eq:M2de}
 \Omega_{\rm wino,dec}h^2 = \left(\frac{M_2}{\rm 3.5\times 10^{-9}~{\rm GeV}}\right)\times Y_{3/2}.
\end{eqnarray}

\begin{figure}[tb]
 \begin{center}
  \includegraphics[width=0.7\linewidth]{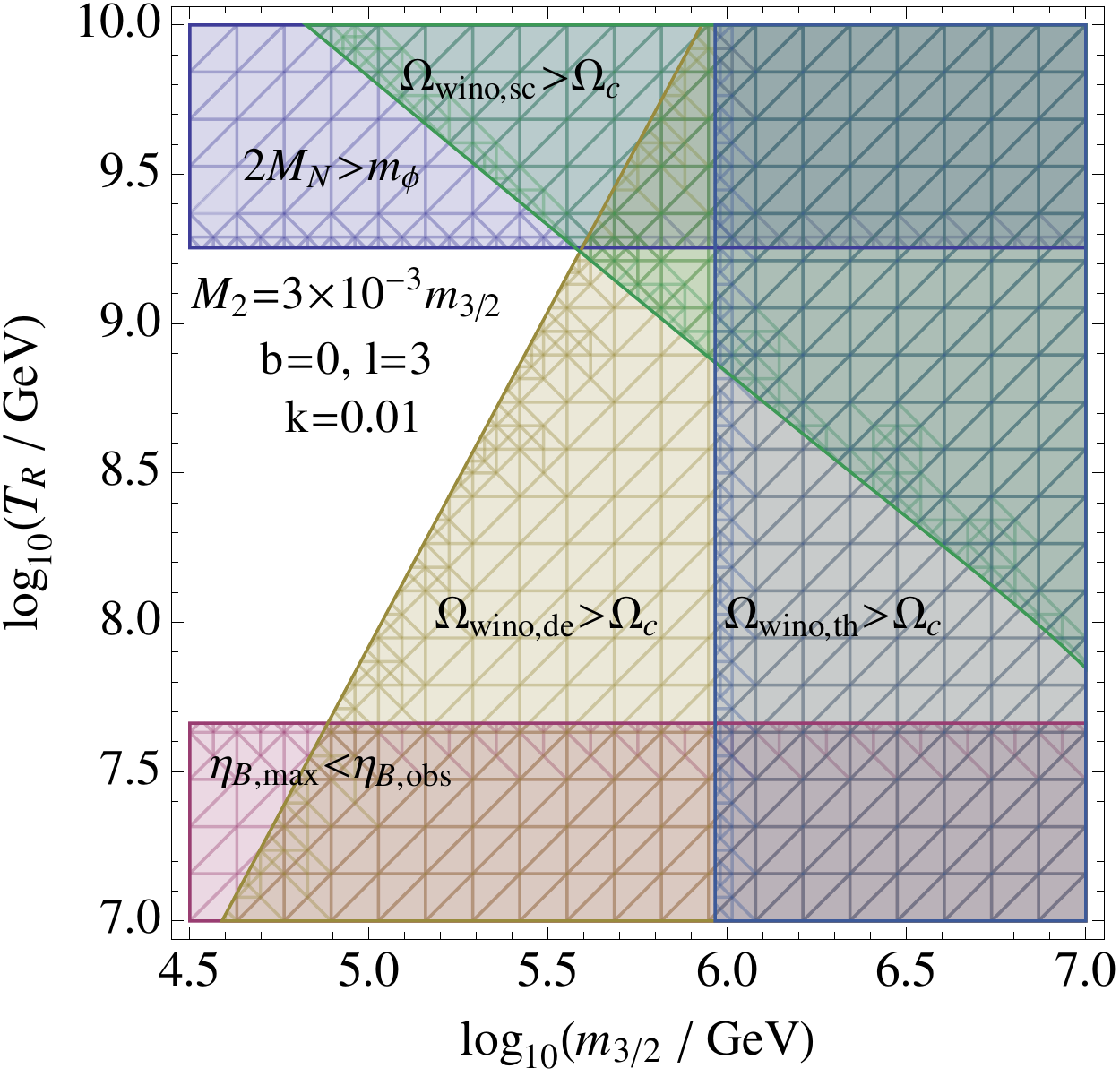}
 \end{center}
\caption{
Constraint on the gravitino
mass and the reheating temperature from the wino
abundance and the successful non-thermal leptogenesis scenario.
Here, we have assumed the wino mass $M_2$ in Eq.~(\ref{eq:AMSB}) with $L=0$.
}
\label{fig:constraint}
\end{figure}

In Fig.~\ref{fig:constraint}, we show constraints on the gravitino
mass and the reheating temperature from the wino
abundance in a $(m_{3/2},T_R)$ plane, which is obtained by Eqs.~(\ref{eq:M2th}), (\ref{eq:M2sc})
and (\ref{eq:M2de}).
Here, we have assumed that the wino mass is given by the purely anomaly
mediated effect, $M_2\simeq 3\times 10^{-3} m_{3/2}$.
The figure shows that the observed dark matter density is mainly explained by 
the non-thermal contributions.
If the coupling constant in the K\"ahler potential, $b$, is close to unity, the
mixing between the SUSY breaking field and the inflaton is suppressed
and hence the contribution from the inflaton decay is small.

We have also shown constraints from the baron asymmetry in the
non-thermal leptogenesis scenario.
The reheating temperature is identified with the one given in Eq.~(\ref{eq:TR}).
In the lowest colored region, the generated baryon asymmetry is smaller
than the observed value even if the $CP$ violation is maximum,
$\delta_{\rm eff}=1$. The result is insensitive to ${\rm tan \beta}$ as
long as ${\rm tan \beta}\gsim 1$.
It can be seen that there is a portion of parameter space in which the baryon
asymmetry as well as the dark matter density in the present universe is
explained.

\section{Summary and discussion}
In this paper, we have investigated a compatibility of the
supersymmetric $R$-breaking new inflation model with the results of the Planck experiment.
We have shown that a lower bound on the
gravitino mass, $m_{3/2}>{\cal O}(100)$ TeV, is obtained 
from the result of the Planck experiment.
We have also shown that the baryon asymmetry as well as the dark matter density in the present universe
can be explained consistently with the $R$-breaking inflation model.

As a final remark, let us interpret the gravitino mass from the landscape
point of view~\cite{Bousso:2000x a,Kachru:2003aw,Susskind:2003kw,Denef:2004ze}.
In the landscape of vacua, it is possible that the gravitino mass is biased to low energy scales in order to
obtain the electroweak scale as naturally as possible. 
In this case, the nature should choose the gravitino mass which saturates the lower bound 
given by Eq.~(\ref{eq:gravitino mass}).
Therefore, the gravitino mass,  $m_{3/2}\simeq100$\,TeV, is a prediction in the $R$-breaking new inflation model
in the landscape point of view.%
\footnote{
If there is a severer bound on $g$ than the
unitarity bound, a larger gravino mass, such as PeV, is predicted from the landscape point
of view.
This arugement may support the explanation of the PeV IceCube neutrino events~\cite{Aartsen:2013bka} by
decaying gravitino
dark matter\cite{Feldstein:2013kka}.
}

It should be cautioned that   there is a hidden parameter in this argument, $k$, 
which has been fixed $k\simeq 0.01$ to account for the observed spectral index. 
From the anthropic point of view, however, there seems no
reason for the spectral index to be close to unity as observed. 
If we allow for a spectral index as large as
$0.8$, for example, then the gravitino mass
is lowered down to,
\begin{eqnarray}
 m_{3/2}\simeq 1.9\times 10^3~{\rm
  GeV}\times\left(\frac{g}{2000}\right)^{-4/5}~~
(n=5,k=0.1,N_e=50),
\end{eqnarray}
which is much smaller than $100$ TeV.

This shows that our landscape argument is self-consistent only
if the parameter $k$ is fixed to be close to $0.01$ by some underlying theory.
If not, the landscape argument predicts that $k\sim 0.1$ and
$m_{3/2}\sim1$ TeV, in which the electroweak scale is obtained much
more naturally than the case with $k\sim 0.01$ and $m_{3/2}\sim 100$
TeV, and the prediction already contradicts with the observed value
of the spectral index.

This situation is similar to anthropic arguments~\cite{Weinberg:1987dv} on the
electroweak scale. It is argued that an electroweak
scale of the one realized in the nature is required for the people to exist in the
universe~\cite{Agrawal:1998xa,Jeltema:1999na}.
There, other parameters other than the Higgs boson mass in the standard model such as the gauge coupling
constants and the Yukawa couplings are fixed to the observed value.
The anthropic prediction on the electroweak scale is viable only if
all such couplings are consider to be fixed by some underlying theory.

\begin{figure}[tb]
 \begin{center}
  \includegraphics[width=0.5\linewidth]{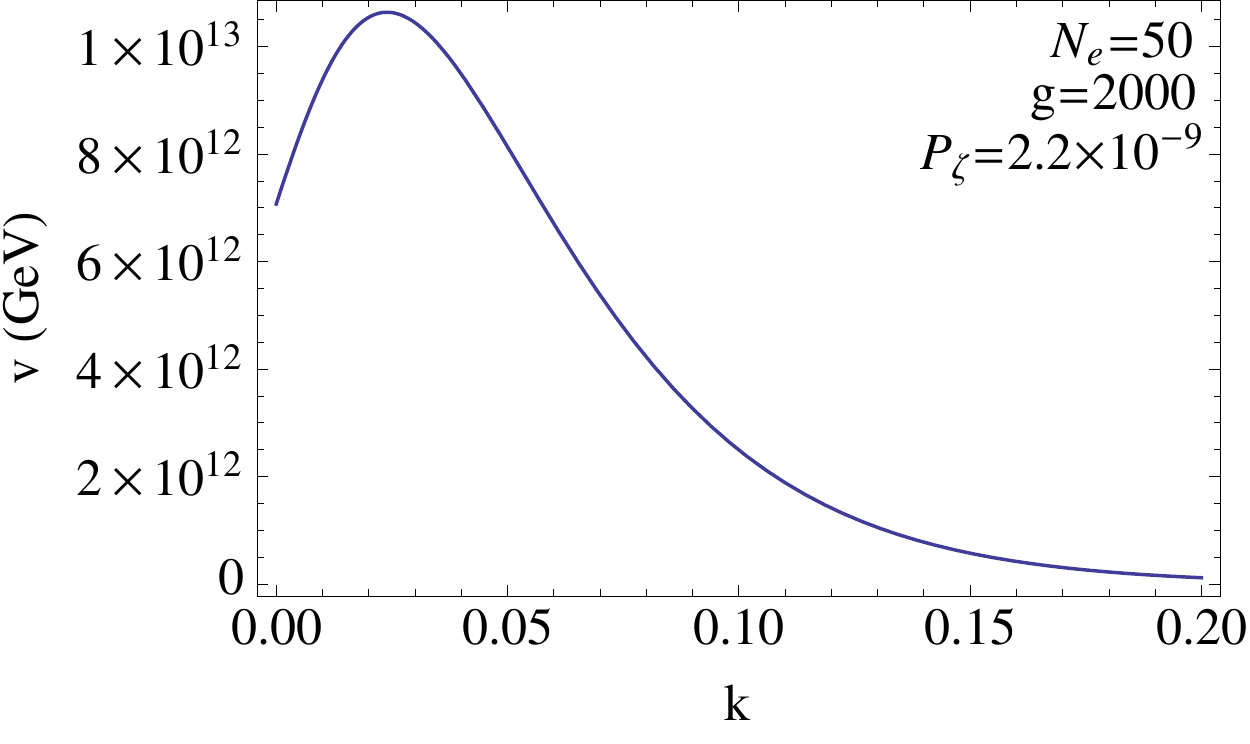}
 \end{center}
\caption{
A line in a $k-v$ space in which ${\cal P}_\zeta =2.2\times 10^{-9}$.
}
\label{fig:kVSv}
\end{figure}

Instead of fixing the parameter $k$, we may move ahead with the
landscape point of view under an additional assumption.
Suppose that the parameter with the positive mass dimension in the
superpotential, $v$, is strongly biased to larger mass scales.
However, $v$ is anthropically required to be sufficiently small in order to generate
a small cosmological perturbation, ${\cal P}_\zeta\sim 10^{-9}$.
Consequently, the maximum $v$ on the hyper-surface 
of the parameter space corresponding to ${\cal P}_\zeta \sim 10^{-9}$
would have been chosen anthropically.
In Fig.~\ref{fig:kVSv}, we show a line in a $k-v$ space in which
${\cal P}_\zeta = 2.2 \times 10^{-9}$.
It can be seen that $k\sim 10^{-2}$, which is
consistent with the observed spectral index, gives the maximum $v$.
Note that the result is insensitive to the parameter $g$.
It is remarkable that a high energy biased $v$ explains the reason why the spectral index $n_s$ is 
not too small such as $0.8$ but close to the observed value, i.e. $n_s\sim0.96$.

\section*{Acknowledgments}
This work is supported by Grant-in-Aid for Scientific research from the
Ministry of Education, Science, Sports, and Culture (MEXT), Japan, No.\ 22244021 (T.T.Y.),
No.\ 24740151 (M.I),  and also by World Premier International Research Center Initiative (WPI Initiative), MEXT, Japan.
 The work of K.H. is supported in part by a JSPS Research Fellowships for Young Scientists.

\end{document}